# Electronic transport in a two-dimensional superlattice engineered via self-assembled nanostructures

Yingjie Zhang[1,2,3]*, Youngseok Kim[2], Matthew J. Gilbert[2] and Nadya Mason[1]*

[1]Department of Physics and Frederick Seitz Materials Research Laboratory, University of Illinois, Urbana, IL, USA

[2]Department of Electrical and Computer Engineering, University of Illinois, Urbana, IL, USA

[3]Beckman Institute for Advanced Science and Technology, University of Illinois, Urbana, IL, USA

*Correspondence to: yjz@illinois.edu, nadya@illinois.edu

**Abstract:** Nanoscience offers a unique opportunity to design modern materials from the bottom up, via low-cost, solution processed assembly of nanoscale building blocks. These systems promise electronic band structure engineering using not only the nanoscale structural modulation, but also the mesoscale spatial patterning, although experimental realization of the latter has been challenging. Here we design and fabricate a new type of artificial solid by stacking graphene on a self-assembled, nearly periodic array of nanospheres, and experimentally observe superlattice miniband effects. We find conductance dips at commensurate fillings of charge carriers per superlattice unit cell, which are key features of minibands that are induced by the quasi-periodic deformation of the graphene lattice. These dips become stronger when the lattice strain is larger. Using a tight-binding model, we simulate the effect of lattice deformation as a parameter affecting the inter-atomic hopping integral, and confirm the superlattice transport behavior. This 2D material-nanoparticle heterostructure enables facile band structure engineering via self-assembly, promising for large area electronics and optoelectronics applications.

**Introduction**

Band structure engineering is key to realizing next-generation electronic and optoelectronic devices. One promising approach toward this goal is superlattice modulation, i.e., engineering long-range periodic patterns to artificially tailor the electronic band structure. This artificial lattice can have a periodicity in the range of 1–50 nm, which is shorter than the electron mean free path and longer than the angstrom-level atomic bond length, inducing the formation of minibands. To date mainly two types of electronic superlattices have been demonstrated: 1) vertically stacked layers with alternating composition, widely used for quantum cascade lasers and infrared photodetectors[1,2]; 2) lateral two-dimensional (2D) Moiré superlattices composed of lattice-aligned graphene/hexagonal boron nitride (hBN) heterostructures[3–6], which are promising for plasmonic modulations and light-harvesting applications[7,8]. However, both systems require lattice-matching conditions, which pose constraints on the choice of materials composition and the extent of device applications.

Self-assembly of solution processed nanoparticles is a low-cost method for nanostructure engineering, which can produce large area close-packed superlattices having overall polycrystalline order and local crystalline domains[9–11]. Despite the high structural quality, superlattice miniband effects have not been experimentally observed in these systems, largely due to the presence of defects on the (semiconducting) nanoparticle surfaces and the weak inter-particle electronic coupling[12–14]. Here we demonstrate a new way to design a 2D superlattice device by integrating solution processed dielectric nanoparticles with 2D materials to form a heterostructure (Fig. 1a), which takes advantage of both the structural versatility of the nanoparticle assemblies and the high mobility of the 2D materials. We observe superlattice miniband conduction in a polycrystalline system where graphene (Gr) is quasi-periodically deformed on top of $SiO_2$ nanospheres (NSs). We find that the size dispersion and imperfect ordering of the NSs induces broadening in the miniband density of states. While this effect is not ideal for applications where sharp states are needed (e.g. lasers & single-color LEDs), it can be beneficial for optoelectronic systems that require broadband modulation of the optical spectrum, such as optical modulators[15], photo-thermal conversion[16,17], infrared sensors[18] and optical communication[19]. While currently we are using CVD grown graphene, this superlattice fabrication method is compatible with solution processed 2D materials[20]. Fully solution processed superlattices can enable low-cost large area applications.

**Results and Discussion**

**Device fabrication and electronic transport**

We use SiO$_2$ nanospheres for heterostructure fabrication because of their insulating nature, clean surface chemistry (terminated with hydroxyl groups and without organic capping molecules), and the ease of comparison with flat SiO$_2$ substrates (same surface chemistry with hydroxyl group termination). The NSs are packed in a polycrystalline structure where each single-crystalline domain consists of tens to hundreds of nanospheres; within each domain the NSs are hexagonal close packed (Fig. 1b, c). Our previous study shows that, after stacking graphene on top of these spheres, Gr bends and stretches around the apex of the NSs, giving rise to a quasi-periodically varying strain pattern[21]. Within the experimentally accessible NS diameter range of 20 – 200 nm (spheres will have irregular shape and large size dispersity if diameter is less than 20 nm), we found that the overall strain in graphene is largest for the smallest sphere diameter[21]. Therefore, here we choose the NSs with 20 nm diameter to induce a strain superlattice in Gr and study the electronic transport. This Gr-NS system shows periodic strain variations in graphene with a peak-to-peak magnitude in the scale of ~2% [21].

We fabricated back-gated 2-terminal devices with Au contacts (Fig. 1a, b), and measured the electronic transport behavior of the Gr-NS systems; we also studied control samples consisting of graphene on flat SiO$_2$/Si substrate (see Methods for details). All the transport results shown here were obtained at 2 Kelvin. Fig. 2a shows the conductance ($G$) as a function of gate voltage ($V_G$) for two control devices (Gr on flat SiO$_2$ substrate, labeled Gr1 and Gr2), which exhibit increasing conductance with gradually decreasing slope at both sides of the Dirac points (DPs). These are characteristic features of graphene-on-SiO$_2$ devices due to the coexistence of long- and short-range scatterers[22–24]. Gr on NS devices (labeled Gr-NS1, 2, 3), in contrast, show emergent conductance dip/kink features in the $G$ vs $V_G$ curves (Fig. 2b). Remarkably, although the kink features occur at different gate voltages for separate devices, all of them are ~15 V away from their DPs. This same separation for different Gr-NS devices indicates that the conductance dips have the same origin. To remove the effect of the (doping induced) DP offsets and the device geometry on the transport characteristics, we plot the conductivity ($\sigma$) as a function of gate-tuned carrier density ($n$) for the same devices in Fig. 2c. Note that conductivity is defined as $\sigma = G\frac{L}{W}$, where $G$ is the conductance, $L$ is the channel length, and $W$ is the channel width (Supplementary Section 1.1 shows the values of $L$ and $W$ and the method to convert gate voltage to $n$). Now it is evident that all three Gr-NS devices show dip features near $n = \pm 1 \times 10^{12}\ cm^{-2}$ (corresponding to a $V_G$ position of ~15 V away from the DP), while the amplitude of the dips varies among different devices and between the electron and hole sides. Note that all the Gr-NS devices show nearly the same minimum conductivity at DPs ($\sigma \cong 2\frac{e^2}{h}$), indicating that all the devices have negligible amounts of random vacancy defects and cracks.

To further quantitatively compare the magnitude of the conductance kinks among different devices, we follow a well-developed protocol to normalize the conductance curves[22], which involves subtracting a series resistance (originating from contact resistance and short-range scatterers) from each curve to restore the linear background, and multiplying the curves by constant factors to normalize them to the same scale (Supplementary Section 1.2). As shown in Fig. 2d, after this normalization procedure, the control sample shows linear conductivity at both sides of the DP. The curves for all the Gr-NS devices perfectly overlap with that of the control sample at $|n| > 1 \times 10^{12}\ cm^{-2}$, and show clear slope-changing features at the characteristic carrier density of about $\pm 1 \times 10^{12}\ cm^{-2}$. The magnitude of the divergence from that of the control device is a measure of the amplitude of the conductance dips of the Gr-NS systems. Note that the electron and hole sides are processed separately for each curve, with different normalization factors, and therefore they do not necessarily align at the DP.

The position of the conductance dips matches with the commensurate filling of 4 electrons per superlattice unit cell (this number originates from the 4-fold spin and valley degeneracies in graphene). This can be seen by taking the lattice constant of the superlattice to be $\lambda$ ~ 21 nm (~20 nm NS diameter and ~1 nm gap between adjacent NSs), and using the supercell area $A = \sqrt{3}\lambda^2/2 \approx 4 \times 10^{-12}\ cm^2$. Then, one electron per supercell corresponds to a carrier density of $n_0 = 1/A = 2.5 \times 10^{11}\ cm^{-2}$, and the dips occur at $n = \pm 4n_0 = \pm 1 \times 10^{12}\ cm^{-2}$. The excellent match of the filling number with the theoretically expected value is strong evidence that the dip features are a consequence of superlattice effects, corresponding to superlattice Dirac points (SDPs). These SDPs occur at the mini-Brillouin-zone boundaries due to the superlattice modulation, where the small density of states leads to dips in conductance[4–6,25].

From the perspective of materials design, the widely studied graphene/hBN system allows a maximum superlattice period of ~14 nm, limited by lattice mismatch conditions of the constituent layers[4–6]. As a result, the SDP position of Gr/hBN systems is at least ~ $2.4 \times 10^{12}$ $cm^{-2}$. Our Gr-NS system offers a route to tune SDP positions in a larger range, including small carrier densities ($< 2 \times 10^{12}$ $cm^{-2}$) that are easily accessible via gate tuning. To further demonstrate this capability, we fabricated and measured a Gr-NS device where the NSs have a diameter of ~50 nm. Transport results reveal an intriguing conductance oscillation with a periodicity of ~$4.5 \times 10^{10}$ $cm^{-2}$, corresponding to 1 electron per supercell (Supplementary Fig. S2c,d), where the supercell now corresponds to the 50 nm NSs. Note that the same 1 e- / supercell oscillation feature (~$2.5 \times 10^{11}$ $cm^{-2}$ periodicity) is also present in the 20 nm system Gr-NS1, in addition to the SDPs at 4 e- / supercell (~$1 \times 10^{12}$ $cm^{-2}$) (Supplementary Fig. S2a,b). Overall, the 4 e- / supercell feature occurs in all of the measured devices having 20 nm spheres, while the 1 e- / supercell feature only occurs in some of the devices. Similar transport behavior has recently been observed in bilayer graphene and trilayer graphene-hBN superlattices, and explained as electron correlation effects[26,27]. While the exact mechanism behind these transport features is still under study and is beyond the scope of this paper, the dip/oscillation features at commensurate filling for Gr-NS systems having two different NS sizes are strong evidence of superlattice effects.

To examine the effect of strain on the superlattice transport properties, it is desirable to deliberately modify the strain of a device while leaving other parameters unaltered, and observe the change in transport features. Previous work shows that temperature cycling can lead to compressive strain in an initially strain-free flake of $Bi_2Se_3$ contacted by electrodes[28]. Here we use the same approach to alter the strain properties of our system. After cooling down from 300 K to 2 K, and then warming up to 300 K again, we find that, via Raman spectroscopy, the doping remains nearly the same while the spatially averaged tensile strain decreases by ~20% (Fig. 3a, Supplementary Section 1.5). The actual change of the nanoscale strain variation amplitude, or the RMS (root mean square) strain, can be much larger than the shift of average strain (Supplementary Section 1.5, Table S2 and Table S3). Meanwhile, transport measurements of a Gr-NS device (Gr-NS2) reveal that the superlattice dip feature becomes much weaker after temperature cycling, compared to the same device before the thermal process (Fig. 3b, c). From the transport results, we also extract the mobility values (proportional to the slope of the curves for $n < -1 \times 10^{12}$ $cm^{-2}$) and find it to have only a minor change (~6% decrease) after temperature cycling, revealing that the modifications of structural defects are small. Therefore we conclude that strain modulation is likely the major factor contributing to superlattice transport features in the Gr-NS devices.

In Fig. 3c we show how the amplitude of the conductance kinks at SDPs can be quantified by the slope changing angle in the normalized conductivity curves ($\theta_1$ and $\theta_2$ for Gr-NS2 before and after temperature cycling, respectively). We observe that $\theta_1 > \theta_2$, which quantitatively confirm the weakening of superlattice modulations. This conductance normalization and "kink angle" method can be generally used to quantify the amplitude of broad conductance dip features in polycrystalline or polydisperse superlattice systems.

In addition to the temperature cycling studies, to decisively prove the strain effect on superlattice transport, we prepared an extra control sample where 20 nm $SiO_2$ nanospheres are assembled on top of graphene on a flat $SiO_2$ substrate. Compared to the graphene-on-NS samples, this control sample is expected to have similar amounts of impurity, doping, and scattering at the graphene-NS interface, but strain will be absent since graphene lies on the flat substrate. Electronic transport results, shown in Supplementary Fig. S3, reveals that a NS-on-graphene device exhibit no conductance dip feature. This result further proves that: 1) the conductance dip features of the Gr-NS devices are due to superlattice effects, instead of random scattering induced by the NSs; 2) strain modulation is the key factor contributing to the superlattice effects in the Gr-NS systems.

**Quantum transport simulations**

While our previous work has revealed periodic strain variations in the Gr-NS systems[21], and now we have experimentally observed the effect of strain on electronic transport in the superlattice (Fig. 3), it is still worth to theoretically examine all the possible factors that can contribute to superlattice transport. One possibility is doping modulation. It is known that random charged impurities are present on the $SiO_2$ surface, which induce p-type doping and charge fluctuations (or charge puddles) in the graphene layer on top[23,24,29–32]. In our Gr-NS samples, the random charged impurities (with a concentration $n_i$) on the NS surfaces, together with the spatial variation of Gr to NS distance, determine the potential distribution of the graphene. From our experimental transport results, we estimate that $n_i \sim 7 \times 10^{11}\ cm^{-2}$ (Supplementary Section 1.6), which agrees with previous reports[23,24,29–31]. This corresponds to ~3 charged impurities per supercell on average, given that the NS diameter is only 20 nm. Considering the random, discrete, and sparse distribution of these impurities, we expect the induced potential modulation in graphene to be mostly random and produce no superlattice effects. This is confirmed by our Coulomb potential and quantum transport simulations (Supplementary Section 2.5).

Another possible superlattice modulation factor is the gate-induced electrostatic potential variation in graphene resulting from its inhomogeneous dielectric environment. However, due to the small height variations of graphene (~2 nm, determined from atomic force microscopy) and the small size of spheres (20 nm) compared to the thickness of the flat $SiO_2$ dielectric layer (300 nm), we expect the gate-tuned carrier density in graphene to be nearly homogeneous. Our electrostatic simulation confirms that, within the experimental range of gate voltage, the gate-induced variation in electrostatic potential is smaller than the random potential fluctuations in Gr caused by the charged impurities on the $SiO_2$ surface (Supplementary Section 1.7). As a result, the electrostatic contribution to superlattice transport should be negligible. This is in agreement with our experimental result shown in Fig. 2b, where we find a strong SDP feature for Gr-NS2 on the hole side occurring near zero $V_G$, revealing that superlattice transport can occur without any electrostatic modulation.

To further quantify the effect of strain variations on carrier transport, we perform tight-binding transport simulations incorporating the experimentally relevant NS size and distribution and Gr strain variation parameters (details in Supplementary Section 2.1 and 2.2). We use the following formula to relate the bond length to the hopping parameter (*t*) in the tight binding Hamiltonian:

$$t + \delta t = t\ exp(-\beta(l/a - 1)), \tag{1}$$

where $a$ and $l$ are the bond length of the unstrained and strained graphene, respectively, and $\beta = 3.37$ is a constant parameter[33]. We chose a range of strain magnitude, and simulate the conductance of the Gr-NS system with both single-crystalline and polycrystalline packing structures (Fig. 4a, b). Note that we directly imported the experimentally imaged NS packing profile for simulation of the polycrystalline system, shown in the inset of Fig. 4b (see also Supplementary Fig. S11). The unstrained graphene (Gr on flat $SiO_2$ substrate) system shows the standard transport features of graphene with no superlattice conductance dips. When strain is added, SDPs appear at the expected energy levels, and become more evident as strain is increased, for both the single-crystalline and polycrystalline systems. The main difference between these two systems is that the polycrystalline devices exhibit broader conductance dips due to the structural disorder. The simulation results on the polycrystalline system are in good agreement with the experimental data in Fig. 2b–d, and imply that the variations in the magnitude of the dip features in different experimental Gr-NS devices can be attributed to differences in strain amplitudes. For example, comparing Fig. 2c, d with Fig. 4b, we can roughly estimate that the RMS strain values of Gr-NS2 and Gr-NS3 are ~2% and ~1% (by comparing the kink angles), respectively. Other factors, such as fluctuations in charge doping, mobility, and NS ordering are either negligible or do not correlate with the change of dip features (Supplementary Section 1.8). The strain variation among different samples is likely due to the device fabrication and cooling down processes.

In summary, we have created a large-area applicable superlattice device by stacking graphene on solution processed assemblies of $SiO_2$ nanospheres. Electronic transport measurements reveal strain-tunable conductance dips due to superlattice miniband effects. Key evidences for superlattice transport in the Gr-NS devices are: 1) conductance dips occur at 4 $e^-$ per supercell for all the Gr-NS devices, which diminish when strain is reduced; 2) control devices of flat Gr and NS-on-flat Gr do not show conductance dips; 3) transport simulations of Gr-NS systems incorporating the actual polycrystalline nanosphere packing structure reveal strain-dependent conductance dips, similar to the experimental results.

We expect that our hybrid heterostructure, combining solution-processed, self-assembled nanoparticle superlattice and atomically thin 2D materials (with superior flexibility compared to bulk[34–37]), will be a powerful framework to design a variety of "artificial solids" that enable band structure engineering via real space patterning. These superlattice devices are cost-effective to manufacture, and are promising for large-area, broadband optoelectronic applications such as optical modulators, infrared sensors, and photothermal converters[15–19].

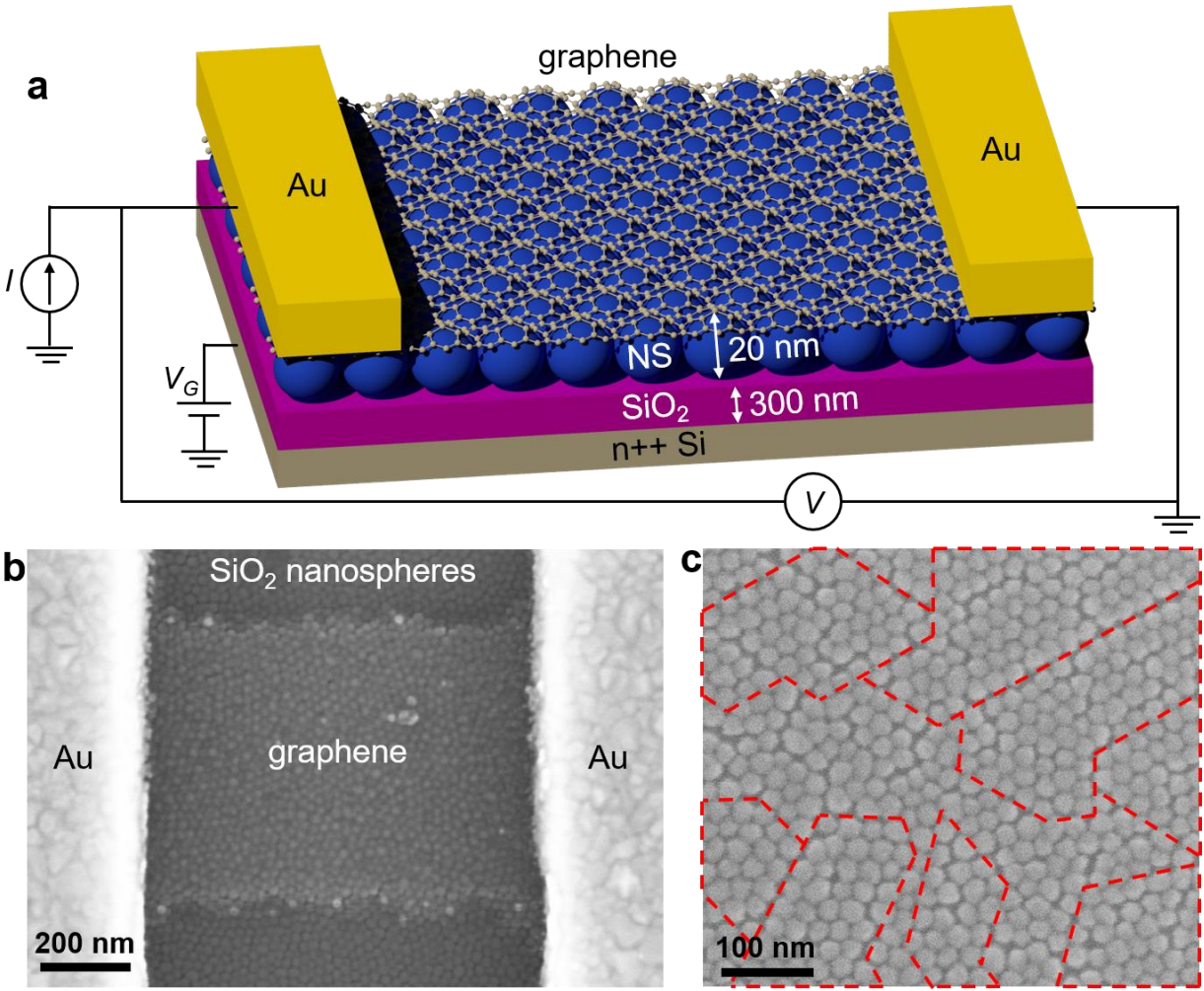

**Fig. 1** Device schematics and images. **a** Schematic of the strain superlattice device structure. SiO$_2$ nanospheres (NSs) with 20 nm diameter are assembled on 300 nm SiO$_2$ / n++ Si substrate. Graphene is stacked on top of the NSs and contacted by 1 nm Cr / 110 nm Au. Channel length and width are between 0.6 – 2.4 $\mu m$ in all the fabricated devices. **b** Scanning electron microscopy (SEM) image of one Gr-NS superlattice device, showing the Au contacts, SiO$_2$ nanospheres, and graphene on nanospheres. **c** High resolution SEM images of a monolayer NS assembly, where each region enclosed by red dashed lines is a single-crystalline domain.

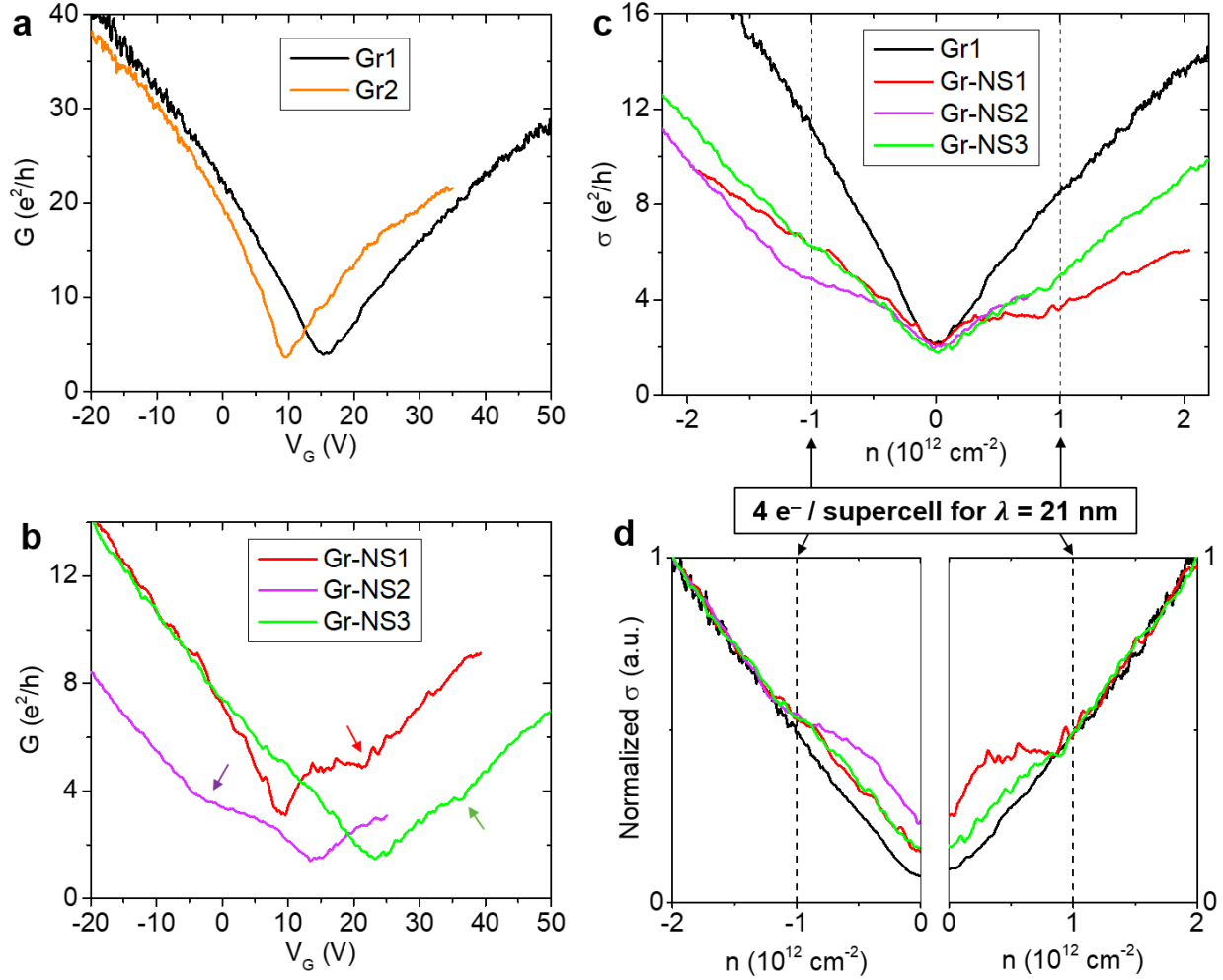

**Fig. 2** Experimental superlattice transport. **a** Conductance ($G$) vs gate voltage ($V_G$) curves for two control devices (Gr on flat $SiO_2$) Gr1 and Gr2. **b** $G$ vs $V_G$ curves for three batches of Gr on 20 nm NS devices as labeled. **c** Conductivity ($\sigma$) vs carrier density ($n$) for a control device (Gr1) and three Gr-NS superlattice devices. The conductivity of Gr1 is multiplied by a factor of 1/2 in order to fit to the same scale as other devices. **d** Normalized conductivity for the same devices shown in (c) (detailed normalization procedures discussed in Supplementary Section 1.2). Vertical dashed lines mark the positions of superlattice Dirac points (SDPs) in all the figures, corresponding to a carrier density of four electrons/holes per supercell. All the transport results shown here were obtained at T = 2 K.

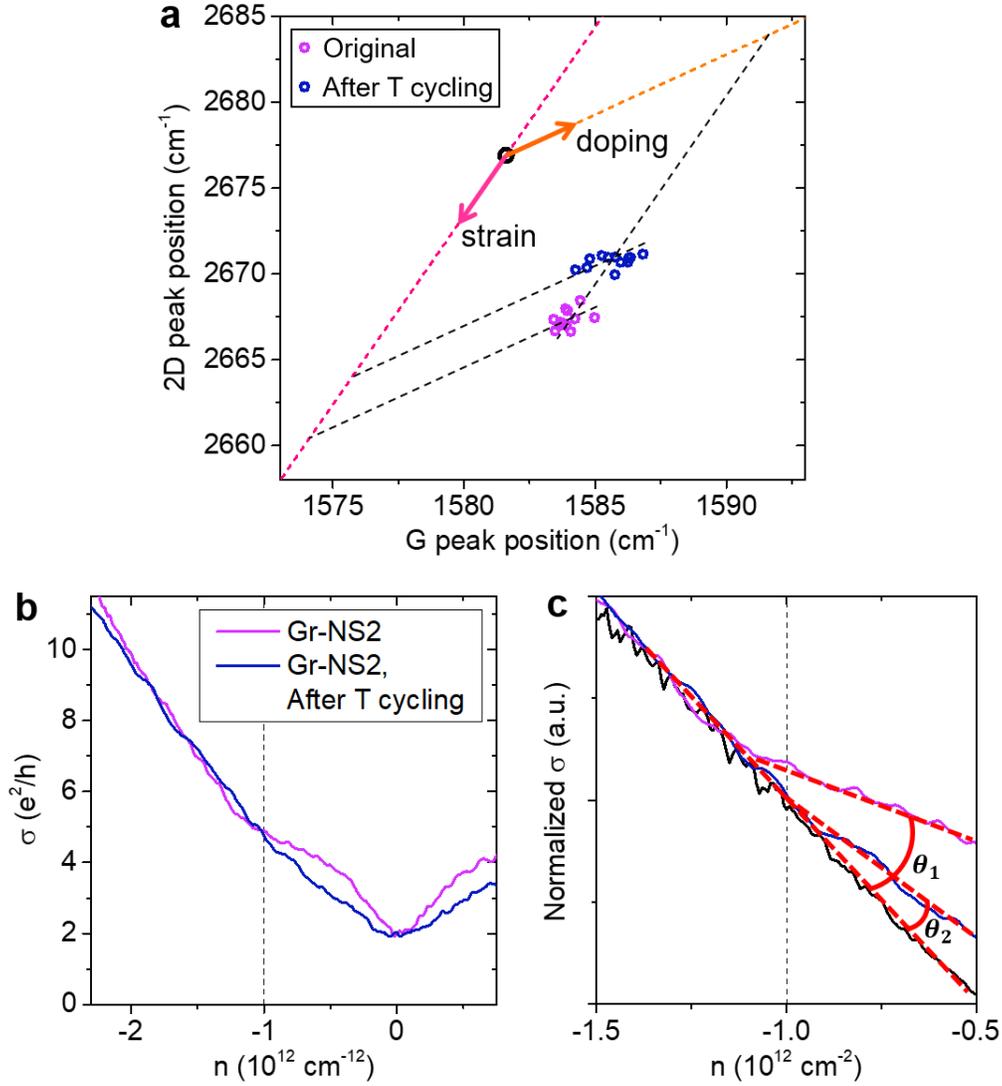

**Fig. 3** Experimental strain manipulation and the effect on superlattice transport. **a** Correlation analysis of the measured Raman G and 2D peak positions, revealing a decrease of spatially averaged strain by 20% ± 4.9% after temperature cycling. In contrast, the average doping value shows negligible change (an increase of 2.3% ± 8.8%). Each point represents a spectrum obtained at an optical pixel with a size of ~0.5 × 0.5 $\mu m^2$ (details in Supplementary Section 1.5). **b** Conductivity vs carrier density for Gr-NS2, after the first cool down to 2 K (purple curve), and after warming up to 300 K then cooling down to 2 K again (dark blue curve). **c** Normalized conductivity for the same devices shown in (b), and the control device Gr1 (black curve, same as that in the left panel of Fig. 2d). The curves are expanded to a carrier density range close to the SDP, and the kink angles of the Gr-NS2 device before and after temperature cycling are labeled as $\theta_1$ and $\theta_2$, respectively.

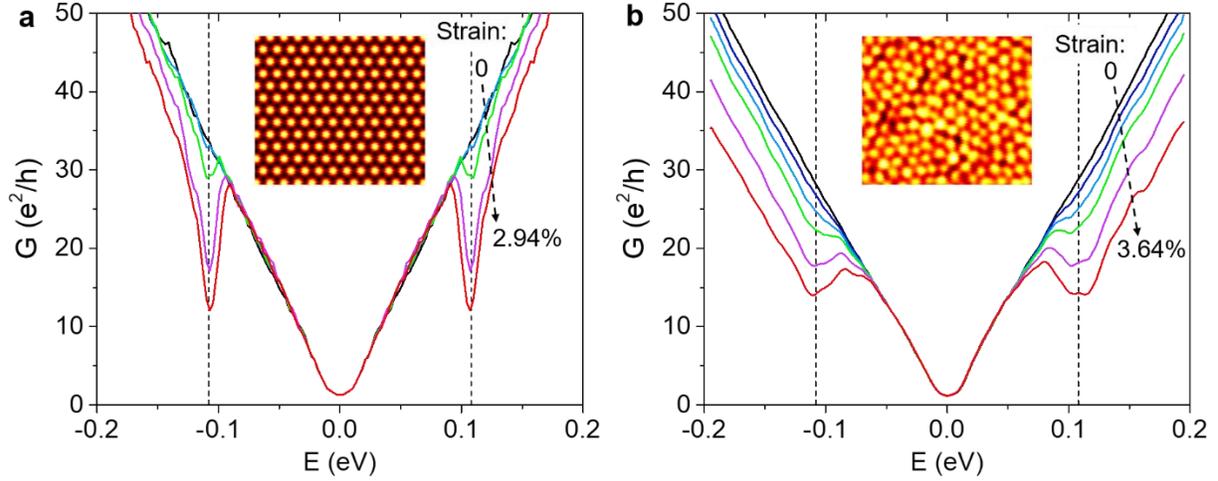

**Fig. 4** Quantum transport simulations. **a** Simulated conductance (*G*) *vs* energy (*E*) curves for graphene on single crystalline NSs (packing structure shown in the inset), with different RMS strain values of 0, 0.28%, 1.11%, 2.02%, 2.94% from top to bottom (black to red) (details shown in Supplementary Section 2.3). **b** Simulated *G vs E* curves for graphene on polycrystalline NSs. The simulated NS packing structure, as shown in the inset, is adopted from our experimental image. From top to bottom (black to red), the curves correspond to RMS strain values of 0, 0.99%, 1.42%, 1.86%, 2.75%, 3.64% (details shown in Supplementary Section 2.4). Vertical dashed lines mark the expected SDP position at $E_{SDP} = \pm \frac{h v_F}{\sqrt{3}\lambda} = \pm 0.11\ eV$, where $h$ is the Planck constant and $v_F = 1 \times 10^6\ m/s$ is the Fermi velocity of graphene[3].

## Materials and Methods

**Sample preparation.** SiO$_2$ nanospheres with 20 nm diameter were purchased from nanoComposix, product number SISN20-10M. The surfaces of these NSs are clean with only hydroxyl groups covering them (no organic capping ligands). The NSs were dispersed in water with a concentration of 5 mg/mL. 300 nm SiO$_2$ / n++ Si substrates were cleaned by acetone, IPA and O$_2$ plasma, followed by spin-coating of the NSs. The sample was then baked on hotplate at 170 °C for 20 min to remove the residual water. The NSs were attached to the SiO$_2$/Si surface with sub-monolayer coverage on the whole substrate, and closely packed monolayer domains in the scale of microns to tens of microns were observed. CVD graphene (purchased from ACS Material) was transferred to the NS substrate using the standard wet-transfer techniques[38], followed by critical point drying to avoid ripping of graphene. We observed both regions of graphene on top of the NS monolayers and regions of Gr on the flat SiO$_2$ substrate on each sample.

**Device fabrication and measurements.** Confocal Raman spectroscopy was performed to identify defect-free areas of graphene for device fabrication. E-beam lithography was performed to define 2-terminal electrodes for both the Gr-NS regions and the flat Gr regions, followed by thermal evaporation of 1 nm Cr / 110 nm Au. The gap between each pair of leads is between 0.6 – 2.4 $\mu m$, and these leads provide structural support for the graphene in the channel region during the following fabrication steps. This is important for the Gr-NS systems where free-standing regions of graphene (between the neighboring NSs) are fragile. For the same reasons, previous high-quality transport measurements of free-standing graphene were performed on 2-terminal devices instead of multi-terminal Hall bar devices[39]. Subsequently, another e-beam lithography step was performed to define etch patterns, and O$_2$ plasma was used to etch graphene into rectangular areas between the 2-terminal leads. After each lithography and lift-off step, the samples were dried in a critical point dryer to avoid structural degradation. The completed devices were wire-bonded and measured in a physical property measurement system (PPMS) under vacuum conditions. An AC current source (10 nA, 17 Hz) was applied to the Au source/drain leads to measure the resistance using a lock-in amplifier, while gate voltage was applied to the Si substrate to tune the carrier density.

**Raman spectroscopy.** Confocal Raman spectroscopy was performed using a Nanophoton Raman 11 system at room temperature, with an excitation wavelength of 532 nm. 100 × objective was used and the laser power was between 0.5–1 mW. All the measured spectra were calibrated using the emission lines of a neon lamp.

**Transport simulations.** A tight-binding model was used to construct the model Hamiltonian for graphene and observables were computed using non-equilibrium Green's function formalism. The form of the Hamiltonian and simulation details are available in Supplementary Section 2.

**Data availability.** The data used in this study are available upon reasonable request from the corresponding author.

**Acknowledgements**

We thank J. W. Lyding for valuable discussions. Y.Z. was supported by a Beckman Institute Postdoctoral Fellowship at the University of Illinois at Urbana-Champaign, with funding provided by the Arnold and Mabel Beckman Foundation. N.M. acknowledge support from the NSF-MRSEC under Award Number DMR-1720633. Y.Z. acknowledge research support from the National Science Foundation under Grant No. ENG-1434147. M.J.G. acknowledge support from the National Science Foundation under Grant No. ECCS-1351871. This work was carried out in part in the Frederick Seitz Materials Research Laboratory Central Facilities and in the Beckman Institute at the University of Illinois.

**Author contributions**

Y.Z. and N.M. conceived the experiment. Y.Z. designed and fabricated devices, and carried out transport measurements. Y.K. and M.J.G. performed the transport simulations. Y.Z., Y.K., M.J.G. and N.M. analyzed the data and wrote the manuscript.

**Competing interests:** The authors declare no competing interests.

# Electronic transport in a two-dimensional superlattice engineered via self-assembled nanostructures

Yingjie Zhang,[1,2,3] Youngseok Kim,[2] Matthew J. Gilbert,[2] and Nadya Mason[1,*]

[1]*Department of Physics and Frederick Seitz Materials Research Laboratory, University of Illinois, Urbana, IL, USA*
[2]*Department of Electrical and Computer Engineering, University of Illinois, Urbana, IL, USA*
[3]*Beckman Institute for Advanced Science and Technology, University of Illinois, Urbana, IL, USA*

## Contents

## 1. Supplementary experimental data and analysis

### 1.1. Carrier density calculation and channel geometry

Dirac points (DPs) are obtained as the gate voltage where minimum conductance occurs ($V_G^D$). The corresponding charge doping density at the DP is $n_D = C_{ox} V_G^D / e$, where $C_{ox}$ is the capacitance of the gate oxide per unit area, and $e$ is the elementary charge. For the control devices, the oxide is 300 nm thick, and $C_{ox} = 1.15 \times 10^{-4}$ F/m$^2$. For the Gr-NS devices, we take the oxide thickness to be 320 nm (300 nm $SiO_2$ film plus 20 nm $SiO_2$ NS), and obtain $C_{ox} = 1.08 \times 10^{-4}$ F/m$^2$. In this calculation we have ignored the height variation of graphene on the NSs ($\sim 2$ nm) which is much smaller than the total oxide thickness. Carrier density is obtained from gate voltage ($V_G$) as $n = C_{ox}(V_G - V_G^D)/e$. The channel geometry and DP position of all the control devices and Gr-NS devices are listed in Table S1.

---

* nadya@illinois.edu

| Device | Gr1 | Gr2 | Gr-NS1 | Gr-NS2 | Gr-NS3 |
|---|---|---|---|---|---|
| $L$ ($\mu m$) | 2.4 | 1.65 | 0.6 | 2.2 | 1.8 |
| $W$ ($\mu m$) | 2.2 | 1.8 | 0.9 | 1.6 | 1.5 |
| $V_G^D$ (V) | 15.5 | 9.6 | 9 | 13.7 | 23 |
| $n_D$ ($10^{12}$ $cm^{-2}$) | 1.11 | 0.69 | 0.61 | 0.92 | 1.55 |

TABLE S1. Channel length (L), channel width (W) and Dirac point position of all the devices.

### 1.2. Subtraction of series resistance

We adopt an empirical model from Morozov *et al.*[1] to analyze the sublinear conductance curves of graphene. The resistance ($R = 1/G$) is taken as a combination of two components $R_L = C_L/n$ (due to long range scattering, where $C_L$ is a constant) and $R_S$ independent of $n$ (due to contact resistance and short range scattering). Therefore we have the fitting equation $1/G = C_L/n + R_S$. For each device, we plot $1/G$ as a function of $1/n$, and do a linear fitting for the high carrier concentration ($|n| > 1 \times 10^{12}$ $cm^{-2}$) part of the curve (where the linear relation is most prominent). This fitting allows us to extract the parameters $C_L$ and $R_S$. The electron and hole sides of the curves are plotted and fitted separately, and different parameters are obtained. Using the fitted parameter $R_S$, we can obtain the series resistance-subtracted conductance as $G_L = 1/(1/G - R_S)$, where $G$ is the raw conductance and $G_L$ is the modified conductance. Fig. S1 show the resistance subtraction process for the control device Gr1 and superlattice device Gr-NS1, as examples.

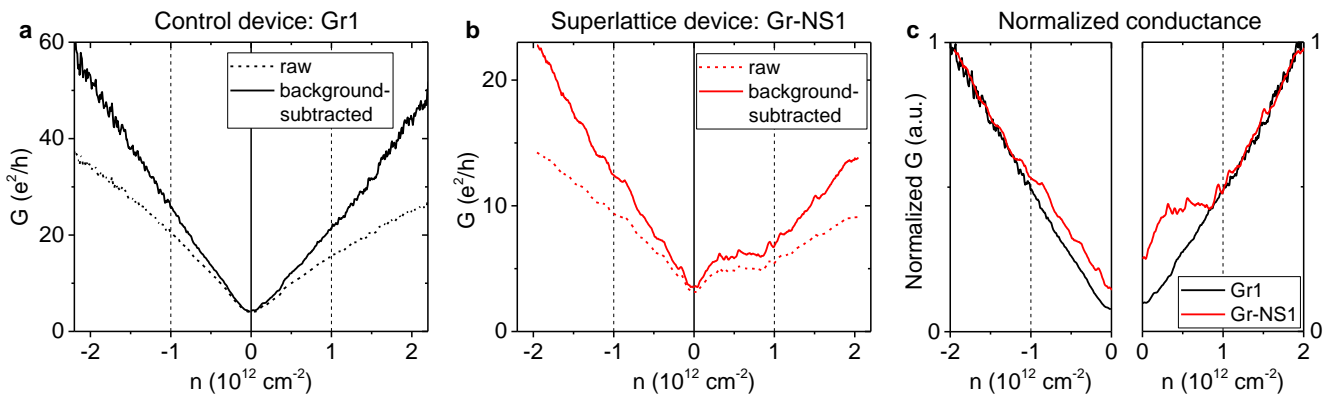

FIG. S1. **Subtraction of series resistance**. (a) and (b) show $G$ (dashed lines) and $G_L$ (solid lines) of Gr1 and Gr-NS1, respectively. We can see that the background-subtracted conductance ($G_L$) of the control device is nearly linear throughout the whole carrier density range (except when very close to the Dirac point, within about $\pm 1 \times 10^{11}$ $cm^{-2}$), while $G_L$ of the superlattice device show nonlinear behaviors and slope changing features at the superlattice Dirac points $n \sim \pm 1 \times 10^{12}$ $cm^{-2}$. To better visualize the differences in these two devices, we simply multiply $G_L$ by a constant factor for each device, so that the two devices show similar conductance values at high carrier density ($|n| \sim \pm 2 \times 10^{12}$ $cm^{-2}$). The normalized conductance curves are shown in (c), where the two devices exhibit overlapping conductance at $|n| > 1 \times 10^{12}$ $cm^{-2}$ and diverging conductance at $|n| < 1 \times 10^{12}$ $cm^{-2}$. Note that the normalized conductivity shown in Fig. 2d is the same as the normalized conductance shown here, since multiplication factors are applied to align different curves together.

### 1.3. Periodicity dependence of superlattice transport

To reveal the effect of different nanosphere sizes on superlatice transport, we prepared NS monolayers where the sphere diameter is $\sim 50$ nm. The shape and monodispersity of the 50 nm NSs are similar to that of the 20 nm NSs. Following the same procedures as the 20 nm systems, we fabricated and measured transport properties of a Gr-50 nm NS device, with results shown in Fig S2c,d. We can see periodic conductance oscillations as a function of carrier density, with a periodicity of $\sim 4.5 \times 10^{10}$ $cm^{-2}$. For the 50 nm system, this corresponds to 1 electron per supercell (assuming 51 nm superlattice period, taking into account the $\sim 1$ nm gaps between adjacent spheres). It is unclear whether there are additional dip features at 4 $e^-$/supercell in this case. For the 20 nm systems, while the 4 $e^-$/supercell conductance dips (at $\sim 1 \times 10^{12}$ cm$^{-2}$) exist in all the measured devices, additional 1 $e^-$/supercell oscillations (at $\sim 2.5 \times 10^{11}$ cm$^{-2}$) also show up in some devices. For example, Gr-NS1 exhibits both small dip features at 1 $e^-$/supercell and broad dips at 4 $e^-$/supercell (Fig. 2b,c and Fig. S2a,b). While we do not fully understand the exact mechanism of the 1 $e^-$/supercell oscillation yet (likely involving degeneracy lifting and/or Coulomb interactions), the commensurate supercell filling features for two different nanosphere sizes are further evidences of superlattice transport.

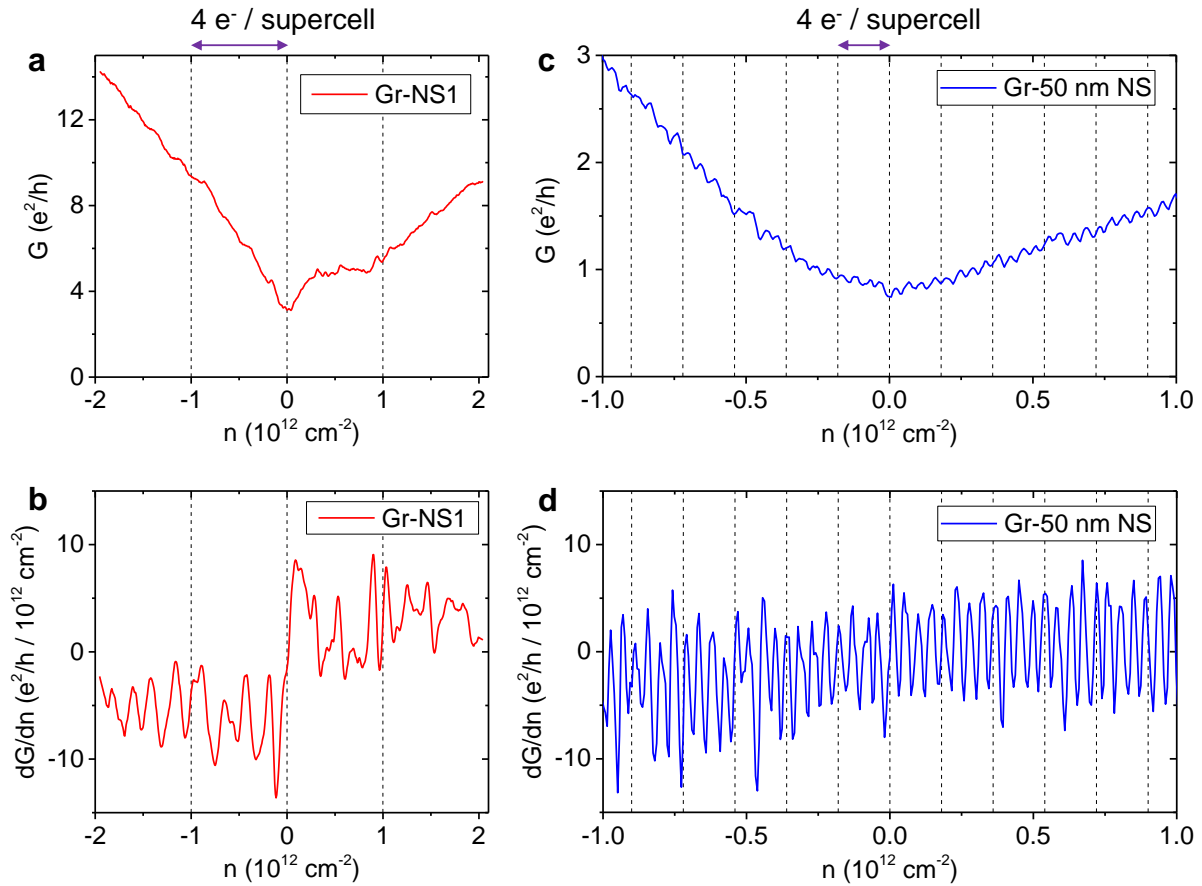

FIG. S2. **Periodicity dependence of superlattice transport**. (a) and (b) show the conductance G and its derivative dG/dn as a function of carrier density for Gr-NS1, where the NS diameter is 20 nm. (c) and (d) show G and dG/dn for a system of graphene on 50 nm NS monolayers. Oscillations are observed in the raw conductance curves (a,c), and the periodicity is more easily identified in the derivative plots (b,d). All the results here were obtained at 2 K.

### 1.4. Control measurements of nanospheres on graphene

We transferred a graphene on a flat SiO$_2$ (300 nm thick) / Si substrate, and assembled SiO$_2$ nanospheres (20 nm in diameter) on top of the graphene. We measured the 2 terminal electronic transport, with results shown in Fig. S3. No conductance dip is observed in this device.

### 1.5. Raman spectroscopy for strain quantification

We adopt the approach of Lee *et al.*[2] to quantify the average strain values of Gr-NS samples. Is it well-known that graphene show two characteristic Raman peaks, the G mode and 2D mode. These two modes are sensitive to both strain and doping effects to different extents. In order to separate these two effects, a correlation analysis of G and 2D modes was proposed and proven to be valid for a variety of graphene systems and heterostructures[2–5].

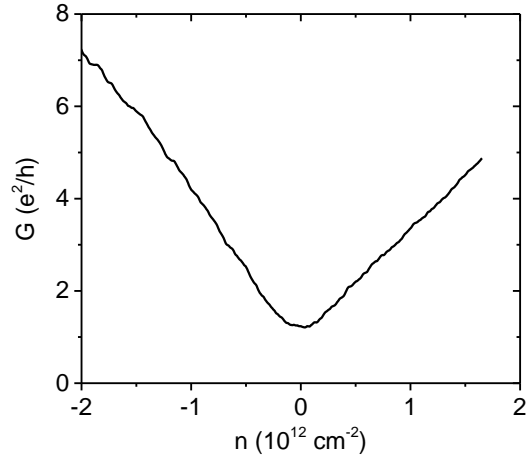

FIG. S3. **Electronic transport of a NS-on-Gr device**. Conductance vs carrier density measured at 2 K.

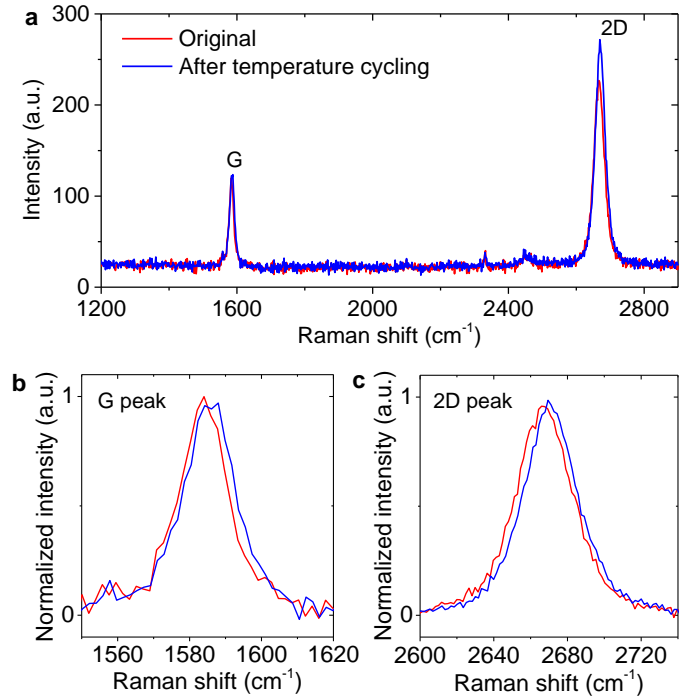

FIG. S4. **Raman spectroscopy of a Gr-NS sample before and after temperature cycling** (from 300 K to 2 K, and back to 300 K). (a) Overall spectrum showing pronounced G and 2D peaks, and the absence of D peak (1350 cm$^{-1}$), confirming that the graphene is single-layer and defect-free (within the detection limit) before and after the T-cycling. (b) and (c) show the expanded G and 2D peaks, where peak shifts show up.

Confocal Raman measurements of Gr-NS were performed at room temperature in ambient conditions. After the first measurement, the sample was transferred to a cryostat to cool down to 2 K in vacuum. After the temperature stabilized, the sample was warmed up to 300 K, and taken out for Raman measurements on the same areas again. The overall spectra before and after the temperature cycling are shown in Fig. S4a, b, c, where spectral shifts are observed. The confocal Raman spectroscopy provides a full spectrum at each optical pixel ($\sim 0.5 \times 0.5$ $\mu$m$^2$), which allows us to fit each spectrum and obtain G-peak and 2D-peak position ($\omega_G$, $\omega_{2D}$) at each pixel.

The collection of ($\omega_G$, $\omega_{2D}$) data points are plotted in the correlation map (Fig. 3a) where $\omega_G$ and $\omega_{2D}$ serve as the X and Y axes, respectively. The intrinsic frequencies of charge-neutral, strain-free graphene are ($\omega_G^0$, $\omega_{2D}^0$) = (1581.6, 2676.9), as shown by the purple hollow circle in Fig. 3a. A doped, unstrained graphene will have ($\omega_G$, $\omega_{2D}$) along the orange dashed line that crosses ($\omega_G^0$, $\omega_{2D}^0$) with a slope of 0.7. The data points of doped graphene can only lie on the right of the charge neutral point, whether the doping is n or p-type. A strained, undoped graphene will have data points along the pink dashed line going across $(\omega_G^0, \omega_{2D}^0)$ with a slope of 2.2. Points below (above) $(\omega_G^0, \omega_{2D}^0)$ along this line correspond to tensile (compressive) strain. If a sample is both doped and strained, the measured $(\omega_G, \omega_{2D})$ point will not be on either the doping or strain lines in the correlation map. In this case a vector decomposition allows us to project the $(\omega_G, \omega_{2D})$ points to the strain and doping lines, and the length of the projected vectors is proportional to the strain or doping values[2]. Following this procedure, we found that the doping remains unchanged after the temperature cycling, while strain decrease from 0.32% to 0.26%. Note that this strain value is measured at room temperature and averaged over micron-scale areas, which is not the same as the nanoscale strain modulation amplitude at low temperature. However, we expect the trend of temperature-cycling induced strain release to be consistent at different temperatures. In fact, besides the cooling-warming cycling, we performed annealing up to 400 °C and cooled down to room temperature, and Raman spectroscopy again showed that strain is reduced after this warming-cooling cycling.

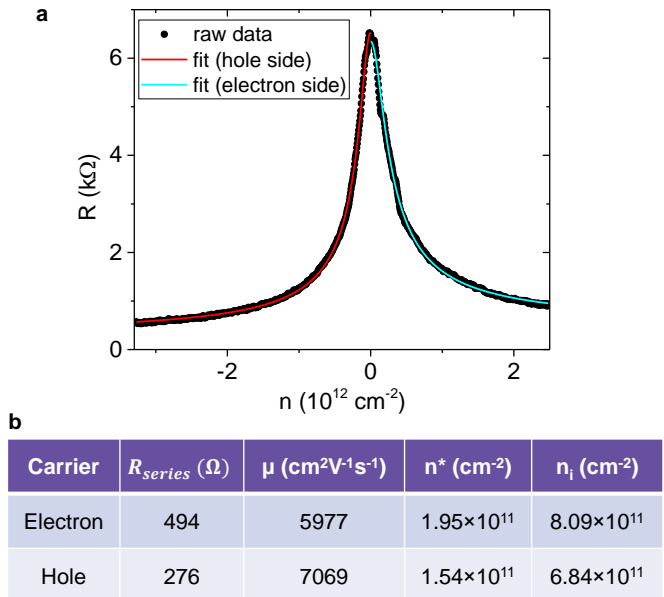

FIG. S5. **Conductance and charged impurity analysis of the control device Gr1**. (a) Raw resistance *vs* carrier density data and the fitted curves using equation (1). (b) Table of the extracted parameters for both electrons and holes.

### 1.6. Charged impurities on $SiO_2$ and residual carrier density in graphene

#### 1.6.1. Charged impurity analysis of the control device

We take the Boltzmann diffusive transport model[6,7] to analyze the charged impurity concentration of the control device Gr1. According to this model, charge carrier transport of graphene on $SiO_2$ is dominantly affected by the random charged impurities (with concentration $n_i$) at the $SiO_2$ surface. These impurities lead to Coulomb scattering that determines the mobility ($\mu$), and induce carrier density fluctuations in graphene. As a result, at Dirac point there is residual carrier density $n^*$ in graphene, known as the "electron-hole puddles"[6-10].

The resistance of graphene can be written as[6]

$$R = R_{\text{series}} + \frac{L}{W}\frac{1}{\mu e\sqrt{n^{*2} + n^2}}, \tag{1}$$

where $L$ and $W$ are the channel length and width in the 2-terminal device, and $R_{\text{series}}$ is the series resistance due to contact resistance and short-range scattering effects. For the control device Gr1, we have $L = 2.4$ $\mu$m and $W = 2.2$ $\mu$m. Using $R_{\text{series}}$, $\mu$, and $n^*$ as the fitting parameters, we perform fitting separately on the electron and hole sides of the resistance curve. As shown in Fig. S5a, this model produces good fits to the experimental data. The extracted mobility is around $6000 - 7000$ cm$^2V^{-1}s^{-1}$, similar with that of the exfoliated graphene device in Samaddar *et al.*[6], confirming the high quality of the CVD graphene we used. The charged impurity concentration of the $SiO_2$ surface is related to the mobility of graphene via[6,7]

$$n_i = 20 \frac{e}{h} \frac{1}{\mu}, \qquad (2)$$

which allows us to determine $n_i$ as shown in Fig. S5b. We found that in our device $n_i$ is around $7 \times 10^{11}$ cm$^{-2}$, in good agreement with previous reports[6–8].

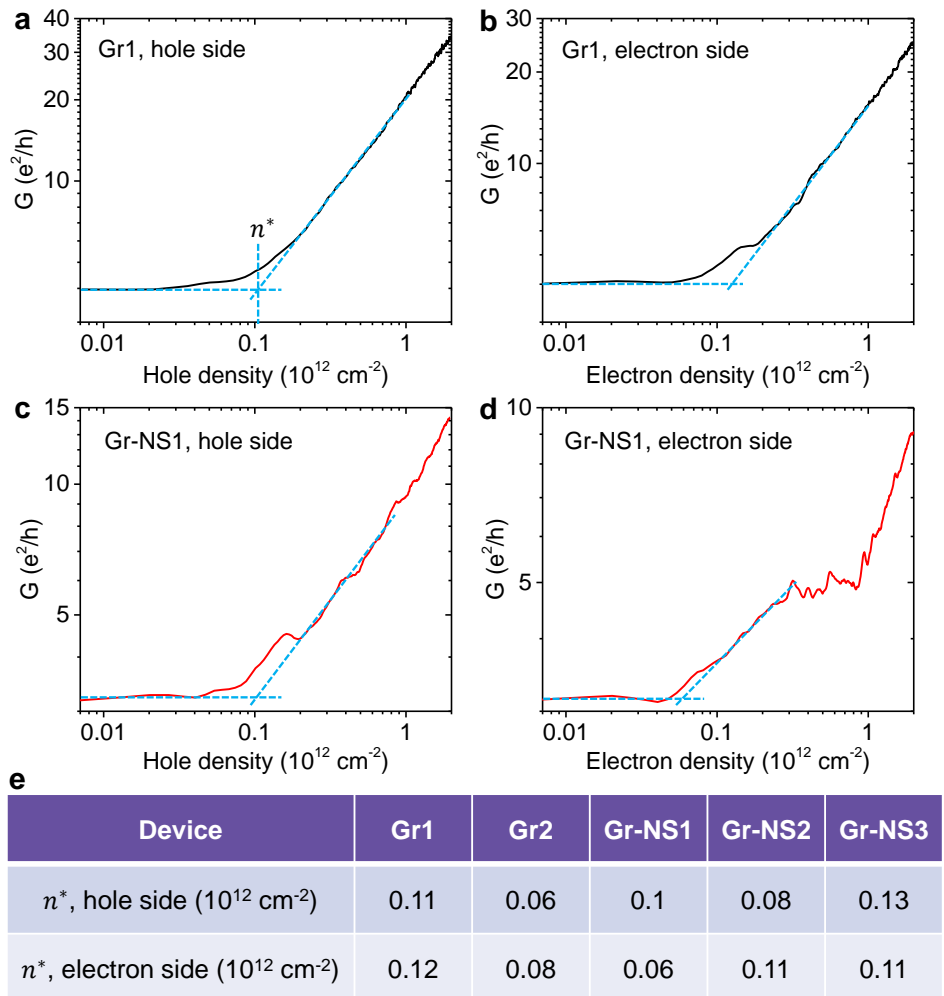

FIG. S6. **Residual carrier density of all the devices**. (a) and (b) show the extrapolation process for $n^*$ of a control device Gr1, at hole and electron side, respectively. (c) and (d) show the same process for a superlattice device Gr-NS1. (e) summarizes the results for all the devices.

### 1.6.2. Residual carrier density analysis of all the devices

While the Boltzmann model described above produces good fit to the transport data of the control device, it is not sufficient to model the transport properties of the Gr-NS devices where the slope of the conductance curves is subject to superlattice modulations. However, we expect that the concentration of the random charged impurities on the $SiO_2$ NS surface to be similar to that of the flat $SiO_2$ surface (both have amorphous structures and the same surface chemistry). And as a result, the residual carrier density in graphene ($n^*$) should be similar. To directly compare the carrier density fluctuation in different systems, we plot the $G$ vs $n$ curves in double-logarithmic scale and extract $n^{*}$[11,12]. The results are summarized in Fig. S6. As we can see, the extracted $n^*$ values for the control device Gr1 (Fig. S6e) are similar to that obtained from the Boltzmann model fitting (Fig. S5b). All of the devices, including the control Gr devices and superlattice Gr-NS devices, show similar $n^*$ in the scale of $1 \times 10^{11}$ cm$^{-2}$. This is evidence that the concentration of random charged impurities ($n_i$) is also similar for all the devices.

### 1.7. Electrostatic profile of the device

To evaluate the uniformness of the gate-induced electrostatic potential in graphene, we perform electrostatic simulations using TCAD software. Fig. S7a describes the two dimensional simulation structure which mimics the fabricated device structure shown in Fig. 1a in the main text. The structure under consideration consists of 300 nm thick SiO$_2$ substrate, SiO$_2$ NSs with 20 nm diameter, and 1 nm thick graphene layer. As conventional TCAD tool has no capability to compute graphene's band structure, we use low-doped silicon as the graphene layer and set dielectric constant $\epsilon_r = 6.9$ in order to mimic the graphene's dielectric environment[13]. The substrate is connected to the gate electrode ($V_G$), and the graphene layer is connected to the source ($V_S$) and drain ($V_D$) electrodes at the leftmost and rightmost edges, respectively. Poisson's equation is solved using Sentaurus Sdevice software and the electrostatic potential within graphene is measured at the center of the 1 nm thick layer while sweeping $V_G$ with $V_S = V_D = 0$ V. We observe a periodic variation in electrostatic potential whose periodicity coincides with that of the NS arrangement. As shown in Fig. S7b, the peak-to-peak variation of the electrostatic potential ($\Delta$V) increases gradually as the gate voltage ramps up, while the slope of the curve (rate of increase) becomes smaller at larger $V_G$. At $V_G = 30$ V, we find $\Delta$V $= 20$ mV.

Typical random potential fluctuations in graphene on SiO$_2$ surfaces (due to random impurity doping effects) are in the scale of $30 - 50$ mV[6,9,14]. Given these disorder effects, the small periodic electrostatic potential variation in our system is thus not likely to induce any measurable superlattice effects by itself. On the other hand, experimentally, we found that superlattice conductance dip can occur even near $V_G = 0$ V where electrostatic modulation is absent (Gr-NS2 in Fig. 2b). Therefore, the observed superlattice transport is likely induced by strain, instead of electrostatic modulations.

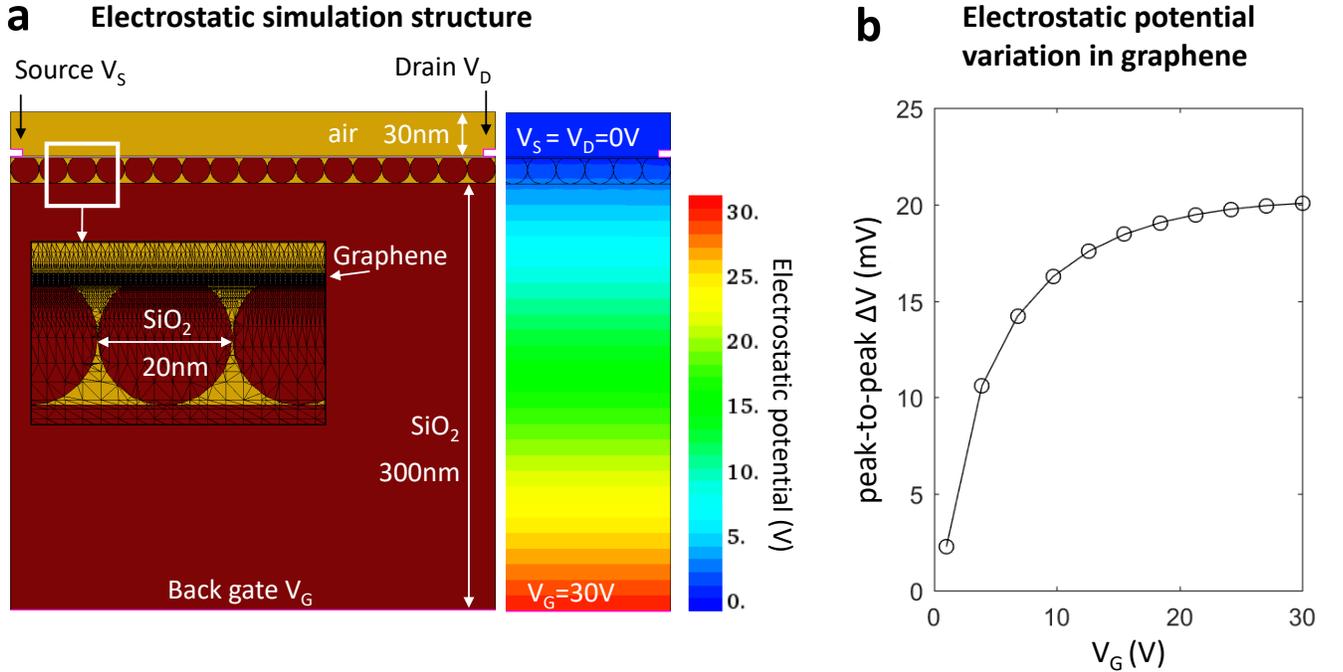

FIG. S7. **Electrostatic profile simulation**. (a) Left: Structure for electrostatic simulation. The inset shows the magnified view of the graphene/SiO$_2$ NSs interface with numerical meshes. Right: the electrostatic potential distribution when $V_G = 30$ V. (b) Peak-to-peak electrostatic potential variation in graphene, plotted as a function of back gate voltage $V_G$.

### 1.8. Comparison of different parameters in Gr-NS devices

#### 1.8.1. Charge doping variations

A direct consequence of the random charged impurities on the $SiO_2$ surface is to dope the graphene and shift the Dirac point position. As shown in Fig. 2b, there are variations in DP positions among different devices, which is typical for graphene-on-$SiO_2$ systems. Given these fluctuations, no systematic difference in charge doping between the control devices and superlattice devices is observed. There is also no obvious correlation between the DP position and the intensity of the superlattice dip features (Fig. 2b, d) in the Gr-NS devices.

#### 1.8.2. Conductivity and mobility

As we can see in Fig. 2c, the overall conductivity slopes are similar among different Gr-NS devices, indicating that their mobility values are also in the same scale. The minimum conductivity at the DPs are around $2e^2/h$ for all the superlattice devices, again proving that these systems have similar levels of structural defects and similar mobility.

#### 1.8.3. Nanosphere packing order

The NS superlattice assembly has a polycrystalline structure with a typical domain size of $100 - 200$ nm. Given the micron-scale size of the graphene channel (Table S1), there are thousands of NSs and tens to hundreds of different crystalline domains in the channel area of each device. Therefore each device should experience similar ensemble average of nanosphere size and polycrystalline packing structure. This is confirmed by SEM images of all the Gr-NS devices.

## 2. Strain and transport simulations

### 2.1. Tight-binding Hamiltonian of graphene

In the Gr-NS system, the substrate surface corrugation induces strain in graphene, manifested as the displacement of the carbon atoms and the change of bond length. As a consequence, the hopping parameter is modulated. In addition, the on-site potential of graphene may also be affected by the variation in the distance between the NS surface and graphene. To include these effects, we write the tight-binding Hamiltonian of graphene as:

$$H = \sum_n \sum_{\alpha=1,2,3} [t_\alpha(\bm{r}_n)c_A^\dagger(\bm{r}_n)c_B(\bm{r}_n + \bm{\delta}_\alpha) + h.c.] \\ + \sum_n V(\bm{r}_n)[c_A^\dagger(\bm{r}_n)c_A(\bm{r}_n) + c_B^\dagger(\bm{r}_n + \bm{\delta}_1)c_B(\bm{r}_n + \bm{\delta}_1)], \tag{3}$$

where $c_A(\bm{r}_n)$ ($c_B(\bm{r}_n)$) is the electron annihilation operator at a site $\bm{r}_n = (x_n, y_n)$ of the sublattice A (B) of the $n^{th}$ unit cell, and $V(\bm{r}_n)$ is on-site potential at $\bm{r}_n$. Note that we do not consider sublattice symmetry breaking in the on-site potential here. In Eq. (3), we define the nearest neighbor vectors

$$\bm{\delta}_1/a = (0,1),\ \bm{\delta}_2/a = (-\sqrt{3},-1)/2,\ \bm{\delta}_3/a = (\sqrt{3},-1)/2, \tag{4}$$

which connect the atoms in the sublattice A and B with an equilibrium inter-atomic distance $a$. Under arbitrary deformations, inter-atomic distance varies spatially and the corresponding nearest neighbor hopping constant changes. In Eq. (3), we define the nearest neighbor hopping constant in the $\bm{\delta}_\alpha$ direction as $t_\alpha(\bm{r}_n) = t + \delta t_\alpha(\bm{r}_n)$ where $t$ is the hopping constant in unstrained graphene and $\delta t_\alpha(\bm{r}_n)$ is a deviation of the hopping constant from $t$.

In this work, we model the graphene using a theoretical artificial graphene lattice[15] whose bond length and hopping parameter are properly scaled to capture the transport properties of the original graphene lattice. For the large scale simulation, we choose the scaling factor $s_f = 10$, resulting the scaled bond length and hopping parameter as $a = s_f a_0$ and $t = t_0/s_f$, respectively, where $a_0 = 1.42$ Å and $t_0 = 2.8$ eV [16].

## 2.2. Strain modulation model

From a microscopic point of view, strain changes the inter-atomic distance of graphene. Then, each hopping parameter follows

$$t + \delta t = t \exp\left(-\beta(l/a - 1)\right), \tag{5}$$

where $l$ and $a$ are bond length of the strained and unstrained graphene, respectively, $t$ is the hopping parameter in equilibrium, and $\beta = 3.37$ is a parameter fitted from the first principle results[17]. We consider two kinds of displacement components that each induces bond length change: an out-of-plane and in-plane displacement. We assume that the out-of-plane displacement is proportional to the height variation of the NS substrate. However, it is not straightforward how to obtain spatial distribution of the in-plane displacement. Typical analytic models of the in-plane strain are based on elastic energy minimization on simplified structures, such as spheres[18–20] or Gaussian bumps[21–23]. Based on these existing results, and the fact that the displacement of atoms is dependent on the protrusion of the substrate structure, we establish a judicious guess on the in-plane displacement profile as

$$\boldsymbol{u} = -f_u \cdot a \nabla h(\boldsymbol{r}), \tag{6}$$

where $h(\boldsymbol{r})$ is the out-of-plane displacement profile, $\boldsymbol{r} = (x, y)$ is the in-plane position of the atoms, and $f_u$ is a unitless in-plane displacement fitting parameter. In Eq. (6), the minus sign is chosen to have a qualitative agreement with the previous simulation results on graphene nanobubbles[24].

In order to check the validity of Eq. (6), we consider the two extreme cases: A small graphene disk lying on top of a sphere (referred to as "graphene disk"), and a graphene sheet much larger than the size of the sphere (referred to as "graphene sheet"). In these two cases, different analytical equations have been developed as a function of radial distance, $r$, from the center of the sphere. Here, we show that our model converges to these two existing models at small and large $r$ limits. First of all, Fig. S8a shows a small disk of graphene film that adheres to the spherical substrate. The resultant stress in radial ($r$) and azimuthal ($\theta$) directions are[18,19]

$$\sigma_{rr}^{(1)}/\Gamma = \alpha(1 - 3(r/W)^2)/16 + 1, \quad \sigma_{\theta\theta}^{(1)}/\Gamma = \alpha(1 - (r/W)^2)/16 + 1, \tag{7}$$

where $W$ is initial radius of the film, $\Gamma$ is adhesion energy per unit area, and $\alpha$ is a confinement parameter determined by the ratio of mechanical and geometric strain contributions[19]. We further utilize the Hookean relationship[19] to obtain strain from stress given in Eq. (7):

$$\varepsilon_{rr}^{(1)} = \frac{\Gamma}{Y}(\sigma_{rr} + v_g \sigma_{\theta\theta}), \quad \varepsilon_{\theta\theta}^{(1)} = \frac{\Gamma}{Y}(\sigma_{\theta\theta} + v_g \sigma_{rr}), \tag{8}$$

where $Y = E_f d$ is the stretching modulus, $E_f$ is Young's modulus, $d$ is thickness, and $v_g$ is the Poisson ratio of the film. In the other scenario, Fig. S8b shows a sheet of graphene that partially adheres to the top of the sphere, while the rest of the region is detached from the sphere. In this case, graphene sheet is treated as a clamped circular membrane with a point load at the center. Then the strain is[23]

$$\varepsilon_{rr}^{(2)}(r) = \frac{3 - v_g}{4}\left(\frac{F}{3\pi^2 E_f^2 d^2}\right)^{1/3}\left(\frac{D}{r}\right)^{2/3}, \quad \varepsilon_{\theta\theta}^{(2)}(r) = \frac{1 - 3v_g}{4}\left(\frac{F}{3\pi^2 E_f^2 d^2}\right)^{1/3}\left(\frac{D}{r}\right)^{2/3}, \tag{9}$$

where $F = E_f(q^3 R_d d)(D/R_d)^3$ is an applied force, $q = 1/(1.05 - 0.15 v_g - 0.16 v_g^2)$, $D$ is the diameter of the NS, and $R_d$ is detachment length. To compare two different models, we adopt parameters[22,23]: $v_g = 0.165$, $d = 0.335$ nm, $E_f = 1$ TPa, $\Gamma = 2.8$ eVnm$^{-2}$, $D = 20$ nm, $R_d = 200$ nm, $\alpha = 5$, and $W = 2.2$ nm.

To examine our in-plane displacement model in Eq. (6), we use the following Gaussian profile,

$$h(x, y) = h_0 \exp\left(-\frac{x^2 + y^2}{\sigma^2}\right), \tag{10}$$

where $h_0$ and $\sigma$ are the height and width of the profile, respectively. By setting the in-plane displacement, $\boldsymbol{u}(x, y) = (u_x(x, y), u_y(x, y))$ and out-of-plane displacement, $h(x, y)$, we have a strain tensor[16]

$$\begin{aligned}
\varepsilon_{xx} &= \frac{\partial u_x}{\partial x} + \frac{1}{2}\left(\frac{\partial h}{\partial x}\right)^2, \\
\varepsilon_{yy} &= \frac{\partial u_y}{\partial y} + \frac{1}{2}\left(\frac{\partial h}{\partial y}\right)^2, \\
\varepsilon_{xy} &= \frac{1}{2}\left(\frac{\partial u_x}{\partial y} + \frac{\partial u_y}{\partial x}\right) + \frac{1}{2}\left(\frac{\partial h}{\partial x}\frac{\partial h}{\partial y}\right).
\end{aligned} \tag{11}$$

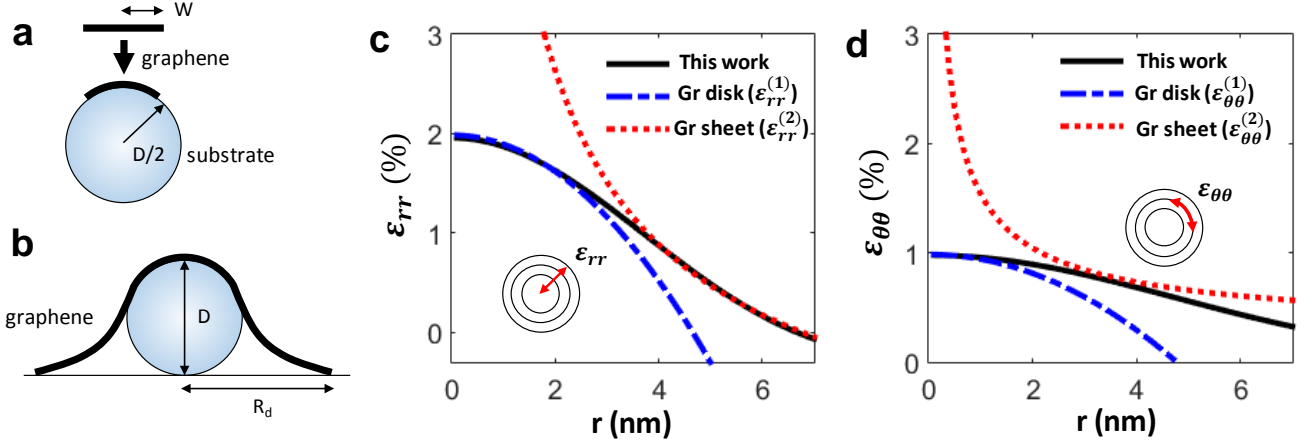

FIG. S8. **Our strain model and comparison with other analytic results.** (a) Schematic of the deformation of a small graphene disk on a spherical substrate. (b) Schematic of the deformation of a large graphene sheet on a spherical substrate. The strain is calculated based on the model for (a), labeled as Gr disk, and (b), labeled as Gr sheet. The resultant radial strain value is plotted in (c) together with the strain calculated using our model in Eq. (6) for a Gaussian profile. Similarly, the azimuthal strain component is plotted in (d).

Solving the strain tensor from Eqs. (6, 10, 11) and transforming the coordinate from cartesian to polar, we obtain

$$\varepsilon_{rr}(r) = 4f_u \frac{ah_0}{\sigma^2}\left(1 - \frac{r^2}{\sigma^2}\right)e^{-r^2/\sigma^2}, \quad \varepsilon_{\theta\theta}(r) = 2f_u \frac{ah_0}{\sigma^2}e^{-r^2/\sigma^2}, \tag{12}$$

where we set $\sigma = D/3$, $h_0 = D/2$. Here, the parameters such as $h_0$ and $\sigma$ are set to make the scale of the Gaussian profile similar with the profile of others in Eqs. (8, 9). In order to present qualitative trend comparison more clearly, we shift the calculated strain profile of $\varepsilon_{rr,\theta\theta}^{(1,2)}$ by adding a constant strain in Fig. S8c, d. Using the fitting parameter $f_u = 0.15$, strain profile of our model in Fig. S8c shows a good agreement with that of the small graphene disk case following parabolic dependence on $r$ when $r$ is small. Note that the spherical strain profile described by $\varepsilon_{rr,\theta\theta}^{(1)}$ is valid only when $r < W = 2.2$ nm, where $W$ is assumed to be the initial radius of the circular graphene disk as shown in Fig. S8a. In this regime, radial and azimuthal strain of graphene on NS described by $\varepsilon_{rr,\theta\theta}^{(2)}$ exhibits diverging behavior which may not be physically relevant. When the radius gets larger, the membrane is detached from the substrate and the strain profile deviates from $\varepsilon_{rr,\theta\theta}^{(1)}$. Instead, our radial strain profile shows $r^{-2/3}$ dependence on $r$, a good match with $\varepsilon_{rr}^{(2)}$, which describes the analytical solution of the radial strain profile in detached region. Therefore, assuming that the Gaussian graphene membrane profile mimics the situation where the membrane partially adheres to the substrate for small $r$ and is detached for large $r$, our model exhibits smooth transition of the radial strain profile from adhered to detached limit. Fig. S8d shows the azimuthal strain profile and our model shows a good agreement with $\varepsilon_{\theta\theta}^{(1)}$ for small $r$ (adhered) limit. Although the azimuthal strain profile deviates from $\varepsilon_{\theta\theta}^{(2)}$ for large $r$ limit, its impact on the total strain value may be minimal as azimuthal strain component is more than 5 times smaller than that of the radial strain in realistic cases[23].

### 2.3. Simulations on single-crystalline superlattices modulated by strain

#### 2.3.1. Height profile and electronic transport

In this section, we show the method for calculating the conductance of graphene with a single-crystalline ordered strain profile. We first construct a Gaussian shape height variation, $h(\boldsymbol{r}) = h_0 \exp(-r^2/2\sigma^2)$ where $h_0 = 3$ nm is the maximum out-of-plane deformation for a Gaussian bump with $\sigma = 10$ nm, and then arrange these bumps in a hexagonal close packed array, with a nearest-neighbor distance of $\lambda = 20$ nm. Fig. S9 shows a periodically modulated height profile of graphene (same as the inset of Fig. 4a). As the Gaussian bump profile overlaps with each other in the periodic arrangement, the height variation becomes $\delta h_0 \sim 0.9$ nm. This height profile is used to obtain the out-of-plane displacement and to compute the in-plane displacement using Eq. (6). Then we compute the change in bond length and hopping parameter using Eq. (5), and we construct the Hamiltonian using Eq. (3). The observables are computed using non-equilibrium Green's function (NEGF) formalism[25]. For example, the density of states are obtained from $DOS(E) = i(G_r - G_r^\dagger)$ where the retarded Greens' function is defined as $G_r = [E - H - \Sigma_L - \Sigma_R]^{-1}$. The left and right contact self energy are defined as $\Sigma_L = \Sigma_R = -it$ using the wide-band limit approximation, and two terminal conductance is obtained from $G = \text{Trace}[\Gamma_L G_r \Gamma_R G_r^\dagger]$, where $\Gamma_{L/R} = 2\text{Im}\{\Sigma_{L/R}\}$.

Fig. 4a shows the conductance obtained from NEGF formalism, where $E_{SDP}$ is plotted as vertical dashed lines. Dip features emerge at $E_{SDP}$ when strain is present, and become more pronounced as the in-plane displacement is increased. To analyze the strain profile in detail, we present the effective strain distribution in section 2.3.2.

### 2.3.2. Strain distribution

Using the height profile in Fig. S9, we obtain the graphene out-of-plane displacement profile. With the fixed out-of-plane displacement of the atoms, we vary the in-plane displacement and calculate the spatial strain distribution using Eq. (11). To evaluate the quantitative behavior of strain for various in-plane displacements, we define the effective strain as

$$\varepsilon_{eff} = \varepsilon_{xx} + \varepsilon_{yy}, \tag{13}$$

which is typically used as a measure of average strain[26,27]. Fig. S10 shows the spatial distribution of the effective strain with increasing in-plane displacement, $f_u = 0, 0.2, 0.4, 1.0$, from left to right column. The resultant strain profile shows local variations, from positive to negative values. In the absence of the in-plane displacement ($f_u = 0$), the strain profile exhibits peak values around the edge of the NSs, similar to the pattern previously shown by STM measurements[28]. As we include the in-plane deformation by introducing non-zero $f_u$, the tensile strain peak is observed near the apex of the NSs, which is in qualitative agreement with molecular dynamics results[24,28,29]. The averaged effective strain values, $\langle \varepsilon_{eff} \rangle$, and the root mean square (RMS) value of the effective strain, $\langle \varepsilon_{eff}^{rms} \rangle$, for various $f_u$ are summarized in the Table S2. Here, $\langle \cdots \rangle$ stands for an average over the simulation area. Although the in-plane deformation substantially increases the effective strain variation up to $\langle \varepsilon_{eff}^{rms} \rangle \sim 4.78\%$ for $f_u = 1.0$, the averaged effective strain is only weakly affected as the spatially averaged in-plane bond modulation is small (Table S2).

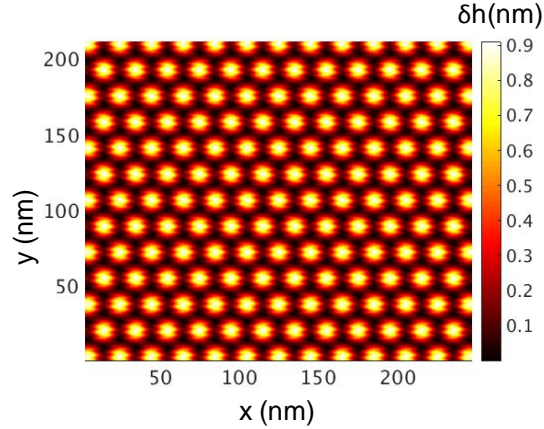

FIG. S9. **Single-crystalline height modulation profile of graphene.** Each bump represents a Gaussian height profile. The system size is $L_x \times L_y = 248 \times 212$ nm$^2$ and carrier transport is along the $\hat{x}$ direction.

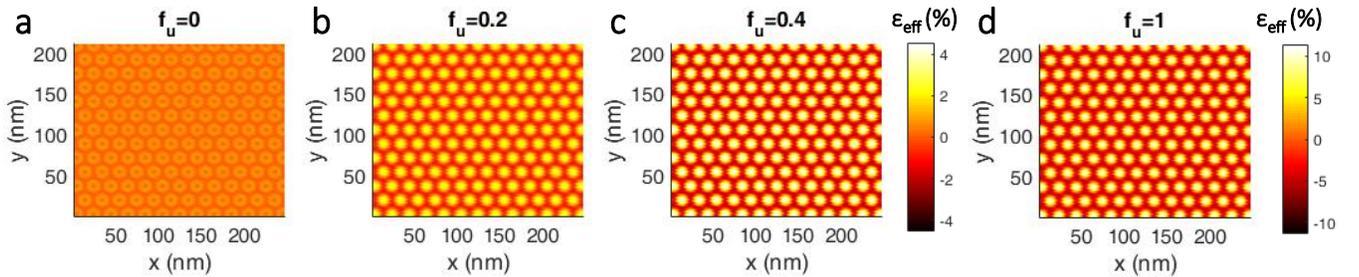

FIG. S10. **Effective strain profile of graphene on NS array for various in-plane displacements.** The effective strain profiles, $\varepsilon_{eff}$, for a structure with only out-of-plane strain ($h_0 = 3$ nm, $f_u = 0$) (a) and for structures with both out-of-plane ($h_0 = 3$ nm) and various in-plane strain ($f_u = 0.2, 0.4, 1.0$) (b-d). The color bar in third column represents $\varepsilon_{eff}$ in % for the the strain distribution plots in (a-c). We use a separate color bar for (d) as the strain variation range is much larger than the other profiles.

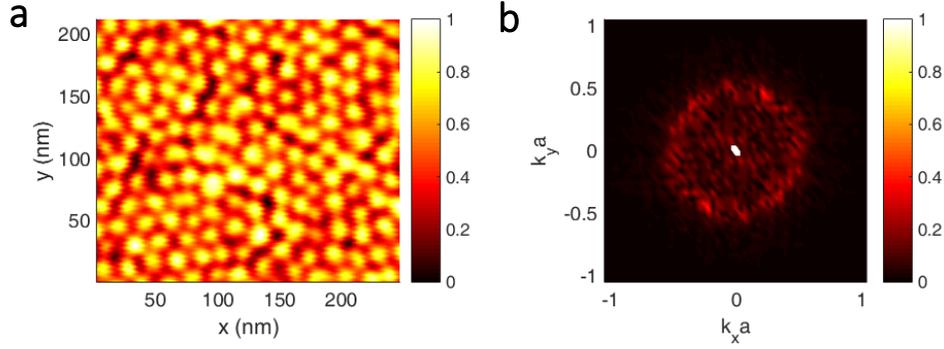

FIG. S11. **A polycrystalline superlattice profile for simulation.** (a) Normalized plot of a polycrystalline ordered superlattice obtained from experimental measurements. The height is normalized and the image is smoothened by Gaussian filter in momentum space to eliminate high frequency noise. (b) Fourier transformed signal of (a). Due to the presence of multiple grain orientations, a circular ring is observed instead of hexagonal point peaks. The amplitude of the signal is normalized.

| $f_u$ | 0 | 0.2 | 0.3 | 0.4 | 0.6 | 0.8 | 1.0 |
|---|---|---|---|---|---|---|---|
| $\langle \varepsilon_{eff} \rangle$ (%) | 0.40 | 0.41 | 0.42 | 0.43 | 0.44 | 0.46 | 0.48 |
| $\langle \varepsilon_{eff}^{rms} \rangle$ (%) | 0.28 | 1.11 | 1.56 | 2.02 | 2.94 | 3.84 | 4.78 |

TABLE S2. Calculated $\langle \varepsilon_{eff} \rangle$ and $\langle \varepsilon_{eff}^{rms} \rangle$ as a function of $f_u$ with $h_0 = 3$ nm for a simulated Gr-NS superlattice system with single-crystalline order (structure shown in Fig. S9).

### 2.4. Simulations on polycrystalline superlattices modulated by strain

In the experimental Gr-NS devices, multiple grain boundaries exist in the NS array within the channel area (the smallest device is $\sim 600 \times 900$ nm$^2$), which can broaden the superlattice dip features. To account for these disorder effects, we use a scanning electron microscopy (SEM) image of the NS assembly with an area of $\sim 200 \times 200$ nm$^2$ that has multiple grain boundaries for simulations. We assume that the geometric height variation in graphene is proportional to the height profile of the NS assembly, which is extracted from the contrast of the SEM image. Fig. S11a shows the obtained height profile of a NS array normalized to $0 \leq p_0(\boldsymbol{r}) \leq 1$, which has a maximum value of 1 on the apex of NSs and minimum value of 0 in the middle of adjacent NSs. Using the height profile in Fig. S11a, we obtain the graphene out-of-plane displacement profile as $h_{sub}(\boldsymbol{r}) = h_0 \cdot p_0(\boldsymbol{r})$, where $h_0$ is the height variation amplitude. Fig. S11b shows a Fourier transformed signal of the NS array profile in Fig. S11a. The signal shows a circular ring instead of hexagonal point peaks due to the misalignment of the NS array lattice vectors between multiple single crystalline domains separated by grain boundaries. To qualitatively estimate the induced strain in the experimentally-relevant Gr on NS array, We use $h_0 = 2$ nm that produces an average effective strain of $\langle \varepsilon_{eff} \rangle \simeq 0.41\%$ or slightly higher, which is close to the experimentally measured (using Raman spectroscopy) effective strain value of $\sim 0.32\%$. The averaged effective strain and RMS effective strain are summarized in Table S3. The results show deviations from the values we obtain from the single-crystalline superlattice (shown in Table S2) due to the difference in the height profile of NSs and the existence of grain boundaries in the polycrystalline system (Fig. S11a).

| $f_u$ | 0 | 0.2 | 0.3 | 0.4 | 0.6 | 0.8 | 1.5 |
|---|---|---|---|---|---|---|---|
| $\langle \varepsilon_{eff} \rangle$ (%) | 0.41 | 0.41 | 0.42 | 0.42 | 0.43 | 0.43 | 0.46 |
| $\langle \varepsilon_{eff}^{rms} \rangle$ (%) | 0.32 | 0.99 | 1.42 | 1.86 | 2.75 | 3.64 | 6.76 |

TABLE S3. Calculated $\langle \varepsilon_{eff} \rangle$ and $\langle \varepsilon_{eff}^{rms} \rangle$ as a function of $f_u$ with $h_0 = 2$ nm for the polycrystalline superlattice shown in Fig. S11a.

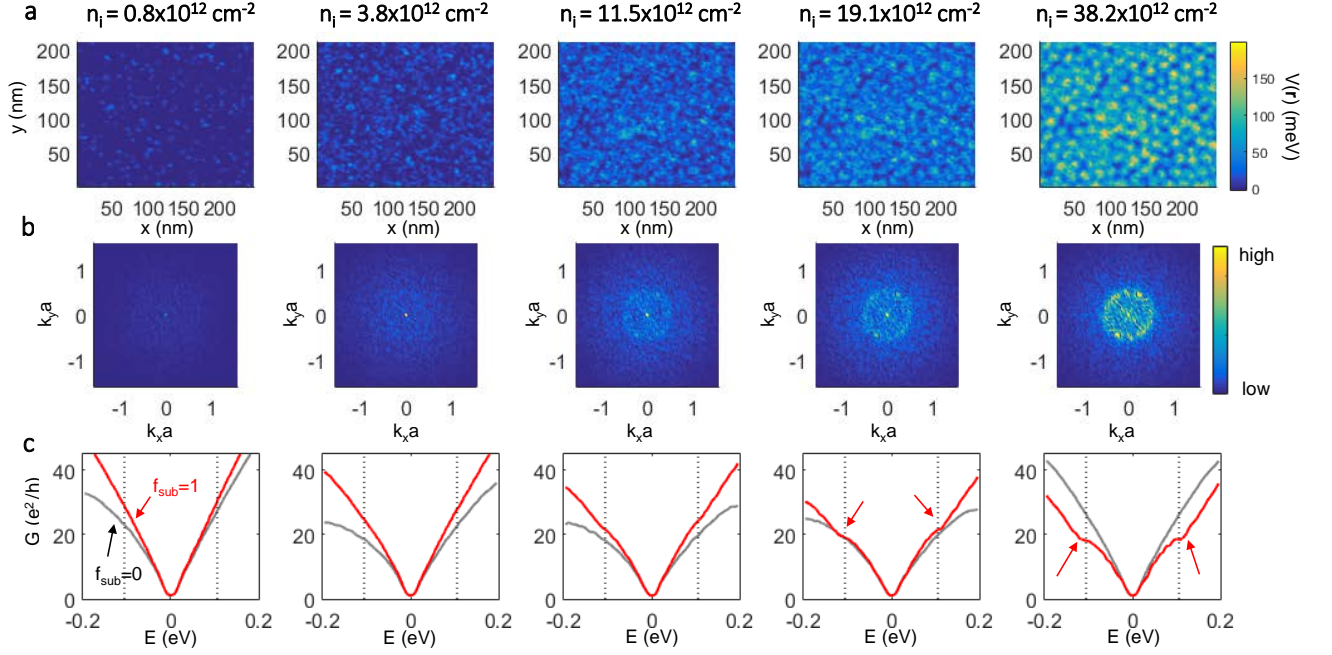

FIG. S12. **Simulated charged impurity distribution and the resultant conductance**. (a) A plot of the potential profile of graphene in the presence of random charged impurities on the NSs. We choose $\delta = 70, 50, 35, 30, 20$ meV and $n_{imp} = 0.021, 0.1, 0.3, 0.5, 1.0$ for the potential profile presented from leftmost to rightmost column. The polycrystalline NSs profile in Fig. S11a is used for height modulation. (b) Fourier transformed signal from the potential profile in (a). (c) A plot of calculated conductance as a function of energy. All the plots are centered at the charge neutrality point and obtained by averaging conductance over 10 different random potential configurations. Potential fluctuation parameters are the same with those of the corresponding columns in (a) and the parameters are chosen to show appreciable degradation in conductance. For all cases, we set $\xi = 1.5$ nm, the system size is $L_x \times L_y = 248 \times 212$ nm$^2$, $N_{latt} = 20000$, and carrier transport occurs along the $\hat{x}$ direction.

### 2.5. Simulations on doping modulation in the superlattice

The major doping source of the graphene on SiO$_2$ is the random charged impurities on the SiO$_2$ surface[9]. The random charged impurities cause the fluctuation in potential, or charge puddles[14]. By fitting its autocorrelation function with the Gaussian profile, we may write the potential[30–32], $V(\boldsymbol{r}_n) = \sum_i^{N_{imp}} U_i \exp\left(-|\boldsymbol{r}_i - \boldsymbol{r}_n|^2/2\xi^2\right)$, where $\boldsymbol{r}_i$ is the $i$th impurity site, $r_n$ is the $n$th atomic site, $U_i$ is the amplitude of the Coulomb potential from the Gaussian scatterers, $N_{imp}$ is total number of impurities, and $\xi$ is the charge puddle size. Both the atomic force microscopy images and cross sectional transmission electron microscopy images (not shown here) reveal that the graphene adheres to the apex of the NSs, and is free-standing in the middle of the adjacent NSs. We assume that the charged impurities are located at the surface of the NSs and, therefore, their contribution is determined by the distance between graphene and the NS surface. We further assume that the Gr to NS distance variation is proportional to the height profile of the NS array, and that the impurity distribution is random. Adopting the height profile of NSs $p(\boldsymbol{r})$ from Fig. S11a, we model the spatial potential distribution as

$$V(\boldsymbol{r}_n) = \sum_i^{N_{imp}} U_i [(1 - f_{sub}) + f_{sub} \cdot p_0(\boldsymbol{r}_i)] \exp\left(-\frac{|\boldsymbol{r}_i - \boldsymbol{r}_n|^2}{2\xi^2}\right), \quad (14)$$

where $f_{sub}$ is a substrate modulation weight parameter varying from 0 (no modulation) to 1 (maximum modulation). To examine the role of the charged impurities in superlattice modulation, we vary the impurity concentration, and calculate the potential profile in graphene and its Fourier transform (Fig. S12a, b), together with the resultant conductance (Fig. S12c). Considering the fact that the random charged impurities result in a net doping in graphene, we set $U_i$ to take a uniform distribution over the interval $[0, 2\delta]$ with a finite averaged impurity potential $\delta$. We first define an impurity concentration in our lattice model as a unitless constant $n_{imp} = N_{imp}/N_{latt}$ where $N_{latt}$ is total number of scaled atomic sites (10 times larger bond length compared to the actual bond length in graphene) and $N_{imp}$ is the total number of impurities. We then vary the impurity concentration from $n_{imp} = 0.021$ (or $n_i \simeq 0.8 \times 10^{12}$ cm$^{-2}$) to $n_{imp} = 1$ (or $n_i \simeq 38.2 \times 10^{12}$ cm$^{-2}$). The potential fluctuation interval, $\delta$, is set to show a noticeable degradation in conductance in order to present a clear change in the conductance plot. When the impurity concentration is in experimentally relevant order of magnitude ($\sim 10^{12}$ cm$^{-2}$), we observe that the major effect of the height modulation is to reduce the average potential fluctuation. When $f_{sub} > 0$, the potential fluctuation at the impurity site with $p(\boldsymbol{r}) < 1$ is reduced and the degradation in conductance due to the Coulomb scattering is recovered. Meanwhile, no signature of superlattice effect is observed due to the fact that only a few impurities are modulated within a length scale of $\lambda \sim 20$ nm. In other words, continuous modulation profile in substrate height does not manifest itself as a modulation in the graphene potential profile due to the sparse sampling by the discrete impurities. For example, there are only $\sim 3$ impurities per supercell for $n_i \simeq 7 \times 10^{11}$ cm$^{-2}$. The leftmost panel of Fig. S12a confirms the sparse distribution of the impurity potential of graphene as isolated peaks and Fig. S12b demonstrates that no evidence of periodicity is observed from Fourier transformed signal of the potential distribution. However, the superlattice effect is observed once the impurity concentration is increased by an order of magnitude. Then the potential profile of graphene exhibits periodicity as shown in the last three columns in Fig. S12b. As a result, SDPs appear in conductance plots, which are marked by red arrows in Fig. S12c. Note that the results on the periodicity of potential profile and superlattice effects are not dependent on the magnitude of the impurity potential fluctuation, $\delta$, as the spatial sampling of the continuous height modulation is solely dependent on the concentration of the impurity. Also, the simulations take $f_{sub} = 1$, the maximum substrate modulation weight parameter, meaning that the part of graphene in the middle of adjacent NSs do not experience charge impurity potential at all. In realistic cases, $f_{sub}$ may be less than 1 and even higher charged impurity concentration is needed to achieve any superlattice modulation.

In this section, we demonstrate that the charged impurities play a negligible role in superlattice effect in graphene unless the impurity density reaches an order of $n_i \simeq 1 \times 10^{13}$ cm$^{-2}$, whereas our devices shows an impurity density of $n_i \simeq 7 \times 10^{11}$ cm$^{-2}$ (average for electrons and holes in Fig. S5b). Therefore, we conclude that charged impurity effects have negligible contributions to superlattice features.

---